# Three-Dimensional Quantum Anomalous Hall Effect in Magnetic Topological Insulator Trilayers of Hundred-Nanometer Thickness

Yi-Fan Zhao[1,4], Ruoxi Zhang[1,4], Zi-Ting Sun[2,4], Ling-Jie Zhou[1], Deyi Zhuo[1], Zi-Jie Yan[1], Hemian Yi[1], Ke Wang[3], Moses H. W. Chan[1], Chao-Xing Liu[1], K. T. Law[2], and Cui-Zu Chang[1]

[1]Department of Physics, The Pennsylvania State University, University Park, PA 16802, USA

[2]Department of Physics, Hong Kong University of Science and Technology, Clear Water Bay, 999077 Hong Kong, China

[3]Materials Research Institute, The Pennsylvania State University, University Park, PA 16802, USA

[4] These authors contributed equally: Yi-Fan Zhao, Ruoxi Zhang, and Zi-Ting Sun

Corresponding authors: cxc955@psu.edu (C.-Z. C.)

**Abstract: Magnetic topological states refer to a class of exotic phases in magnetic materials with their non-trivial topological property determined by magnetic spin configurations. An example of such states is the quantum anomalous Hall (QAH) state, which is a zero magnetic field manifestation of the quantum Hall effect[1]. Current research in this direction focuses on QAH insulators with a thickness of less than 10 nm (Refs. [1-9]). The thick QAH insulators in the three-dimensional (3D) regime are limited, largely due to inevitable bulk carriers being introduced in thick magnetic TI samples. Here, we employ molecular beam epitaxy (MBE) to synthesize magnetic TI trilayers with a thickness of up to ~106 nm. We find these samples exhibit well-quantized Hall resistance and vanishing longitudinal resistance at zero magnetic field. By varying the magnetic dopants, gate voltages, temperature, and external magnetic fields, we examine the properties of these thick QAH insulators and demonstrate the robustness of the 3D QAH effect. The realization of the well-quantized 3D QAH effect**

**indicates that the nonchiral side surface states of our thick magnetic TI trilayers are gapped and thus do not affect the QAH quantization. The 3D QAH insulators of hundred-nanometer thickness provide a promising platform for the exploration of fundamental physics, including axion physics and image magnetic monopole[10], and the advancement of electronic and spintronic devices to circumvent Moore's law.**

**Main text:** Quantum-based technologies promise revolutionary advances in computation, communication, and metrology, but success hinges on the creation of a material platform in which a quantum state can be made robust against environmental decoherence yet simultaneously amendable to control and ease in measurement. The quantum anomalous Hall (QAH) insulator with dissipation-free chiral edge current in magnetic topological insulator (TI) thin films/heterostructures provides a promising platform for this purpose[1]. The realization of the QAH effect requires an interplay between topology and magnetism[1,11,12]. Through the careful introduction of magnetic ion dopants into TI films in a molecular beam epitaxy (MBE) chamber, the QAH effect has been successfully realized in Cr- and/or V-doped $(Bi, Sb)_2Te_3$ films (Refs. [2,13]). To date, most of the experimental efforts on QAH insulators focus on the magnetic TI films/heterostructures with a thickness of less than 10 nm (Refs.[3,4,6-9,14-21]), which are in the two-dimensional (2D) or near the 2D-to-3D boundary regimes [7,21,22]. The 3D QAH insulators are limited, due to inevitable bulk carriers being introduced in MBE-grown thick magnetic TI samples.

Different theoretical frameworks have been used to understand the QAH effect in magnetic TI films/heterostructures within 2D and 3D regimes[1]. In the 2D limit, a hybridization gap ($\Delta_H$) between the top and bottom surface states appears [22-24] and thus makes the TI film a quantum spin Hall (QSH) insulator with helical edge states [25-27]. By introducing transition metal ions into TI, the ferromagnetic order eliminates one spin channel of helical edge states, and the QSH insulator transforms into a QAH insulator, provided that the magnetization exchange gap $\Delta_m$ is greater than $\Delta_H$ (Ref. [11,28]). This framework becomes inappropriate when the TI film becomes thick and undergoes a transition into the 3D regime, i.e., the formation of Dirac surface states with $\Delta_H=0$ (Refs. [22-24]). For 3D TI films, the manifestation of the QAH effect is linked to the presence of the axion term $\theta \frac{e^2}{2\pi h} \boldsymbol{E} \cdot \boldsymbol{B}$ in the modified Maxwell Lagrangian [29], where $\boldsymbol{E}$ and $\boldsymbol{B}$ are the electric and magnetic fields inside TI, and $\theta$ is a pseudo-scalar with its value to be $\pi$ in TI and 0 in vacuum (or in a trivial insulator). When a magnetic exchange gap is induced to break time-reversal symmetry (TRS), the variation of $\theta$ value from $\pi$ to 0 across the surface leads to a contribution of $\sim e^2/2h$ Hall conductance at both the top and bottom surfaces, resulting in the QAH effect with the total Hall conductance $\sigma_{xy}$ quantized at $\sim e^2/h$ (Ref. [12]).

In magnetic TI thin films/heterostructures with the QAH state, the criterion typically employed to differentiate between 2D and 3D limits is the occurrence of hybridization between the top and bottom surfaces[1]. In electrical transport measurements, the emergence of the hybridization gap can lead to a zero Hall conductance plateau, which has been used to differentiate between 2D and 3D QAH effects[7,11]. However, the appearance of the zero Hall conductance plateau can originate from either the QAH insulators with substantial magnetic disorder [17,30-32] or the antiparallel magnetization alignment between the top and bottom surface layers (i.e., the axion insulator phase)[33-35]. These two possibilities specifically impact the connection between the 2D QAH effect and the appearance of the zero Hall conductance plateau in transport measurements. Consequently, the 2D-to-3D crossover thickness for the QAH insulators is usually influenced by the quality of the magnetically doped TI films/heterostructures and ranges from several to a dozen quintuple layers (QL) (Refs.[1,21]).

In this work, we employ MBE to grow thick magnetic TI trilayers with different magnetic dopants in their surface layers, specifically 3 QL Cr-doped (Bi, Sb)$_2$Te$_3$/$m$ QL (Bi, Sb)$_2$Te$_3$/3 QL Cr-doped (Bi, Sb)$_2$Te$_3$ (designated as Cr-$m$) and 3 QL Cr/V-co-doped (Bi, Sb)$_2$Te$_3$/$m$ QL (Bi, Sb)$_2$Te$_3$/3 QL Cr/V-co-doped (Bi, Sb)$_2$Te$_3$ (designated as Cr/V-$m$). In the magnetic TI sandwich, the time-reversal symmetry is broken specifically on the top and bottom surfaces while being maintained within the interior of the sample. This facilitates the realization of the QAH effect[1,8]. Moreover, the significant reduction of defects and disorders in the middle undoped TI layers results in its bulk insulating properties. This allows us to substantially increase the thickness of the QAH samples. We find that the QAH effect persists in the sample with a thickness of up to 106 QL (i.e., $m$ = 100). By varying magnetic dopants, gate voltages, temperature, and external magnetic fields, we examine the properties of these thick QAH insulators and demonstrate the persistence and robustness of the 3D QAH effect. Compared to the QAH trilayers doped only with Cr, the QAH trilayers with Cr/V-co-doping show a wider quantization plateau and a slightly larger thermal activation gap. In addition, we find that the peak values of the longitudinal resistance near the coercive field can be adjusted by ~8.8 times by varying solely the bottom gate while the sample remains within the quantization regime. Through theoretical calculations, we show that the quantum confinement effect of the top and bottom magnetic TI induces a gap on the side surface of the sample. This explains why the QAH effect persists in samples of hundred-nanometer thickness. The thick 3D QAH insulators in the 3D regime provide a good platform for the explorations of axion physics [e.g., the topological magnetoelectric (TME) effect][12], image magnetic monopole [10], and higher-order TI phase with chiral hinge states [36,37].

All magnetic TI sandwiches are grown on ~0.5 mm thick heat-treated SrTiO$_3$(111) substrates in a commercial MBE chamber (Omicron Lab10) with a base vacuum better than ~2×10$^{-10}$ mbar (Methods). The Bi/Sb ratio in each layer is optimized to tune the chemical potential of the trilayer sample near the charge neutral point. Electrical transport measurements are conducted using both a Physical Property Measurement System (PPMS, Quantum Design DynalCool, 1.7K, 9T) for $T \geq$ 1.7 K and a dilution refrigerator (Leiden Cryogenics, 10mK, 9T) for $T < 1.7$K. The magnetic field is applied perpendicular to the sample plane. The high magnetic field measurements are carried out with a superconducting magnet (SCM4-28T, 50mK) and a hybrid 45T magnet (300mK) in the National High Magnetic Field Laboratory (NHMFL) at Tallahassee. More details about the MBE growth, sample characterizations, and transport measurements can be found in Methods.

We first focus on the Cr-100 trilayer sample (Device #1). Figure 1a shows the cross-sectional scanning transmission electron microscopy (STEM) image and the corresponding energy dispersive spectroscopy (EDS) mapping of the Cr element near the bottom surface layer. The STEM image shows that the thicknesses of the top Cr-doped TI, middle undoped TI, and bottom Cr-doped TI layers are ~3 QL, ~100 QL, and ~3 QL, respectively. The total thickness of the Cr-100 trilayer is ~106 QL (Fig. 1a). The ordered layer structure is shown in the STEM image (Fig. 1a). The streaky "1 × 1" reflection high-energy electron diffraction (RHEED) patterns (Supplementary Fig. 1), and the sharp X-ray diffraction (XRD) peaks (Supplementary Fig. 2), indicate the high crystalline quality of our thick magnetic TI trilayers, which might be a pivotal factor for achieving the 3D QAH effect therein (Fig. 1b). Next, we perform electrical transport measurements on Device #1 (Cr-100). At the charge neutral point $V_g=V_g^0$, the Cr-100 sample exhibits the well-quantized $C = 1$ QAH effect at $T=25$ mK (Fig. 1b). The zero magnetic field Hall resistance $\rho_{yx}(0)$ and longitudinal resistance $\rho_{xx}(0)$ are ~0.985 $h/e^2$ and ~0.003 $h/e^2$ (~81.6 Ω), respectively. The coercive field $\mu_0H_c$ is ~0.21 T. All these values are comparable to the $m=4$ trilayer sample with a similar Cr-doping concentration in our prior studies [38].

Besides the QAH trilayers doped with only Cr, we also synthesize thick magnetic TI trilayers co-doped with Cr and V ions. Here we take the Cr/V-70 trilayer (Device #2) as an example (Supplementary Fig. 4a). The Cr/V-70 trilayer also shows the well-quantized $C=1$ QAH effect at $V_g=V_g^0$ and $T=25$mK. The value of $\rho_{yx}(0)$ is found to be ~1.00 $h/e^2$, concomitant with $\rho_{xx}$ ~0.0007 $h/e^2$ (18.6Ω) (Supplementary Fig. 4b). To further validate the QAH effect in both Devices #1 and #2, we measure $\mu_0H$ dependence of $\rho_{yx}$ and $\rho_{xx}$ at different $V_g$ and plot $\rho_{yx}(0)$ and $\rho_{xx}(0)$ as a function of ($V_g$-$V_g^0$) (Fig. 1c and Supplementary Fig. 4c). Here the value of $V_g^0$ is determined when $\rho_{yx}(0)$ is maximized. For Device #1 (Cr-100), both $\rho_{yx}(0)$ and $\rho_{xx}(0)$ exhibit a quantized plateau [with $\rho_{yx}(0) \geq 0.985$ $h/e^2$ and $\rho_{xx}(0) \leq 0.005$ $h/e^2$] for -8 V $\leq$ ($V_g$-$V_g^0$) $\leq$ +20 V, confirming the realization of the QAH effect in Device #1 (Cr-100) (Fig. 1c and Supplementary Fig. 5a). Compared to Device #1 (Cr-100), Device #2 (Cr/V-70) shows a wider quantization plateau between -13 V $\leq$ ($V_g$-$V_g^0$) $\leq$ +90 V (Supplementary Figs. 3c and 5b). The smaller residual $\rho_{xx}(0)$ and broader quantized plateau in the Cr/V-co-doped sample might be a result of the more uniform magnetization [9] and the possible weaker gating effect as a result of the pinning effect of V ions [13].

The observation of the well-quantized QAH effect in thick magnetic TI sandwiches implies that the side surfaces of the sample are insulating, which will be discussed later. To further examine the properties of the chiral edge transport and the gapped nonchiral side surface states in our thick QAH samples, we perform three-terminal transport measurements on Device #1 (Cr-100) (Fig. 2a). We find that both local and non-local three-terminal resistances exhibit perfect quantization when the sample magnetization is well aligned (Fig. 2b). This observation suggests the dissipationless chiral edge transport exists in QAH insulators with a thickness of ~106 QL. Next, we perform electrical transport measurements on Device #2 (Cr/V-70) under $\mu_0H$ up to ~28 T. Figures 2c and 2d show the $\mu_0H$ dependence of $\rho_{xx}$ and $\rho_{yx}$ in Device #2 (Cr/V-70). At $V_g=V_g^0$, Device #2 shows perfect quantization for 0 T ≤ $\mu_0 H$ ≤+28 T at $T$=50 mK (Fig. 2d), indicating the absence of nonchiral edge/surface channels[39]. At $(V_g-V_g^0)$ = -30 V (i.e., the sample is hole-doped), $\rho_{yx}(0)$ ~0.946$h/e^2$ and $\rho_{xx}(0)$ ~0.136$h/e^2$, respectively. As $\mu_0 H$ increases, $\rho_{yx}$ gradually increases and saturates at ~0.990 $h/e^2$ near $\mu_0 H$~20 T, and $\rho_{xx}$ correspondingly decreases to 0.005$h/e^2$, showing an evolution to the $C$=1 Chern insulator state as $\mu_0 H$ localizes the dissipation channels (Fig. 2c). This observation agrees well with that observed in 5 QL uniformly Cr-doped TI films [2] but different from 10 QL uniformly Cr-doped TI films [4] and 10 QL magnetic TI sandwiches[39]. In the latter two cases[4,39], $\rho_{xx}$ initially increases and then decreases as $\mu_0 H$ increases, and a large value of $\rho_{xx}$ persists even at $\mu_0 H$ ~ 10T. This might indicate that the deviation from the QAH state under high magnetic field is attributed to the chemical potential of the sample not being inside the magnetic exchange gap [4,39]. We note that for Device #2 (Cr/V-70), both $\rho_{yx}$ and $\rho_{xx}$ evolve smoothly with $\mu_0 H$, showing no phase transition except for the magnetization switching at $\mu_0 H_c$, which signifies that the $C$=1 QAH state is robust, without the involvement of Landau physics for 0 T ≤ $\mu_0 H$ ≤ +28 T. We further demonstrate the robustness of the $C$=1 QAH state by the $(V_g-V_g^0)$ dependence of $\rho_{yx}(0)$ and $\rho_{xx}(0)$ under different $\mu_0 H$ (Figs. 2e and 2f).

Besides the Cr/V-70 sample, we also perform electrical transport measurements on the Cr/V-100 and Cr-100 samples under high magnetic fields. For the Cr/V-100 sample, the well-quantized QAH effect appears for 0 T ≤ $\mu_0 H$ ≤ +28 T and $V_g = V_g^0$ = -13V. However, for $V_g$ =+200 V (i.e., electron-doped), the sample shows a well-quantized QAH effect at $\mu_0 H$ =0 T, $\rho_{yx}(0)$ ~0.977 $h/e^2$ and $\rho_{xx}(0)$ ~0.014 $h/e^2$. With increasing $\mu_0 H$, both $\rho_{yx}$ and $\rho_{xx}$ gradually deviate from the quantized values. At $\mu_0 H$ =28 T, $\rho_{yx}$~0.915 $h/e^2$ and $\rho_{xx}$~0.153 $h/e^2$(Supplementary Fig. 6). This observation further confirms that the deviation from the QAH state under high magnetic field is not an intrinsic property of the QAH insulators. For the Cr-100 sample, we find that this sample exhibits a negative magnetoresistance for $(V_g-V_g^0)<0$ but a positive magnetoresistance for $(V_g-V_g^0)>0$ (Supplementary Fig. 7). This unexpected deviation from the QAH state with increasing $\mu_0H$ for $(V_g-V_g^0)>0$ could potentially arise from either the enhanced backscattering of the nonchiral surface states or the manifestation of weak antilocalization in thick middle undoped TI layer under higher $\mu_0H$. More studies are needed to pinpoint the physical origin of this phenomenon.

To further investigate the properties of our thick QAH trilayers, we perform more magneto-transport measurements on Device #1(Cr-100) and Device #2 (Cr/V-70) at different temperatures (Supplementary Figs. 8 and 9) and plot $\rho_{yx}(0)$ and $\rho_{xx}(0)$ as a function of $T$ (Fig. 3a). For $T<300$ mK, the transport behavior is predominantly dictated by variable range hopping, which arises from the strong localization effect caused by local disorders [40]. For 300 mK $< T <$ 1 K, the transport behavior is primarily determined by the effective activation gap $E_a$, which can be described using the Arrhenius equation: $\sigma_{xx}=\sigma_{xx}^0 e^{-\frac{E_a}{k_BT}}$. To estimate $E_a$, we plot $\sigma_{xx}(0)$ as a function $1/T$ on a log scale and perform linear fitting for the five sandwiches with $m>50$ (Fig. 3b and Supplementary Figs. 10 and 11). The estimated values of $E_a$ are ~47.7 μeV, ~55.1 μeV, ~56.2 μeV, ~79.1 μeV, and ~83.5 μeV for the Cr-50, Cr-60, and Cr-100 (Device #1), Cr/V-70 (Device #2), and Cr/V-80 trilayers, respectively (Fig. 3c). These $E_a$ values in our thick QAH trilayers are comparable to those estimated in thin QAH samples in prior studies [5,9,16,20,40,41]. This also indicates that the side surfaces are gapped in our thick QAH trilayers. A relatively larger gap size is consistent with the better quantization and the wider quantization plateau in Cr/V-co-doped QAH trilayers, presumably due to their more uniformly distributed ferromagnetic order [9]. For $T>1$ K, the bulk carriers appear and the sample deviates from the QAH state with increasing temperature. The critical temperatures of the QAH state, which are defined as the temperature at which $\rho_{yx}(0)/\rho_{xx}(0)=1$ are ~4.0 K and ~5.1

K for Device #1(Cr-100) and Device #2 (Cr/V-70), respectively. Both are also comparable to the thin QAH trilayers in our prior studies [38,42].

Next, we highlight one unique property of our thick Cr/V-co-doped QAH trilayers. We find that in Device #2 (Cr/V-70), the peak value of $\rho_{xx}$ near $\mu_0H_c$ (denoted as $\rho_{xx,max}$) can be adjusted by altering $V_g$ while the sample still resides in the well-quantized QAH regime. For -22 V< ($V_g$-$V_g^0$) < -2 V, $\rho_{xx,max}$ ~ 1.0 $h/e^2$ (Figs. 4a and 4e). For ($V_g$-$V_g^0$) > 0 V, $\rho_{xx,max}$ steadily increases to ~8.8$h/e^2$ at ($V_g$-$V_g^0$) = +88 V (Figs. 4b to 4e). We note that Device #2 (Cr/V-70) shows a well-quantized QAH effect for -22 V< ($V_g$-$V_g^0$) < +88V (Figs. 4a to 4d). The $V_g$-modulated $\rho_{xx,max}$ can be understood through the following scenario. For the bottom 3 QL Cr/V-doped TI layer, the $\mu_0H_c$ usually varies with ($V_g$-$V_g^0$) and is typically larger by introducing hole carriers [i.e., ($V_g$-$V_g^0$) < 0] [13]. In these thick Cr/V-co-doped QAH trilayers, a single bottom gate is insufficient for effectively tuning the chemical potential of both bottom and top surfaces simultaneously. This can make the $\mu_0H_c$ values of the top and bottom 3 QL Cr/V-doped TI layers different. Therefore, the top and bottom magnetic TI layers can form antiparallel magnetization alignment when $\mu_0H$ is swept between two $\mu_0H_c$s, which drives the thick trilayer from the QAH state towards an axion insulator-like state [33-35]. However, for ($V_g$-$V_g^0$) <0 V, the sample deviates from the QAH state much faster as the magnetic exchange gap is closer to the bulk valance bands [1,5,43,44]. As a result, the bulk carriers smear the $\rho_{xx}$ peak so that we do not observe a large $\rho_{xx,max}$ for ($V_g$-$V_g^0$) <0 V. We next convert $\rho_{xx}$ and $\rho_{yx}$ into longitudinal conductance $\sigma_{xx}$ and Hall conductance $\sigma_{xy}$ (Figs. 4f to 4j). With increasing ($V_g$-$V_g^0$), a zero $\sigma_{xy}$ plateau appears near $\mu_0H_c$, concomitant with a $\sigma_{xx}$ dip feature. Therefore, the thick Cr/V-co-doped QAH trilayer offers a promising platform for investigating a gate-tunable quantum phase transition between the QAH and axion insulator states.

In the following, we discuss the physical origins behind the persistence of the QAH effect in thick magnetic TI trilayers. As noted above, the realization of the 3D QAH effect in thick magnetic TI films/heterostructure needs to eliminate the conduction through the bulk and along the side surfaces[1]. In our magnetic TI trilayers, the reliable MBE growth of the middle undoped (Bi, Sb)$_2$Te$_3$ layer enables us to effectively eliminate bulk conduction. To clarify the influence of the nonchiral side surface states, we perform theoretical calculations based on a magnetic TI trilayer model (Methods). Because of the confinement of the magnetic exchange gap in both top and bottom surfaces, the nonchiral side surface states inherited from the TI are also gapped, with a minimal gap $\delta$ (Fig. 5a). Our theoretical calculations show that the dependency of the side surface gap $\delta$ on the top and bottom surfaces exchange gap $2M$ and the sample thickness $m$ has two features (Fig. 5b). First, for a fixed $m$, there is a critical value in $M$, marked as $M_c$. For $M < M_c$, $\delta$ is larger than $2M$, the side surface states are governed by $\delta$ merging into the bulk states, and we cannot calculate $\delta$. For $M > M_c$, $\delta$ is smaller than $2M$, thus the side surface states governed by $\delta$ can be seen. Moreover, if $M \gg M_c$, $\delta$ will saturate. Second, with increasing $m$, $M_c$ monotonically decreases. For instance, when $2M = 7$ meV, the sample with $m \leq 100$ displays a side surface gap $\delta$ larger than the exchange gap $2M$ on the top and bottom surfaces as $2M_c$ are all larger than 7 meV (Fig. 5b). Therefore, this nonchiral side surface gap does not influence the QAH quantization behavior, as demonstrated by the temperature-dependent measurements (Fig. 3). However, with a further increase in $m$, the value of $\delta$ becomes less than $2M$ and vanishes in thick samples, eventually leading to the disappearance of the QAH effect (Figs. 5c to 5e and Supplementary Fig. 12). We note that the QAH effect will persist as long as the chemical potential is tuned into the nonchiral side surface gap $\delta$, until $\delta$ becomes smaller than the thermal activation energy ($k_B T$) (Figs. 5c to 5e).

Finally, we estimate the upper critical thickness for the QAH state, where the conducting gapless side surfaces emerge. In thick QAH samples, the total longitudinal conductance is $\sigma_{xx} = \sigma_{xx}^{t\&b} + \sigma_{xx}^{side}$. Where $\sigma_{xx}^{t\&b}$ is the thermally activated conductance from both the top and bottom surface states, which phenomenologically takes $\sigma_{xx}^{t\&b} = K^{t\&b}\exp(-2M/k_BT)$ (Refs. [5,39]) The conductance from the side surface states $\sigma_{xx}^{side}$ has contribution only from nonchiral rather than chiral edge states, this contribution can be expressed as $\sigma_{xx}^{side} = K^{side}\exp(-\delta/k_BT)$. $\sigma_{xx}$ is dominated by the smaller value between $2M$ and $\delta$. Note that $2M$ remains independent of thickness, whereas $\delta$ scales as ~$1/m$, like in 1D confinement problems. Therefore, in the large thickness limit, $\sigma_{xx}$ is primarily determined by $\delta$. When the side surface gap $\delta$ is comparable to $2k_BT$, chiral edge states mix with other high-energy nonchiral side surface states through thermal activation, leading to the suppression of the QAH state. For the Cr-100 (Device#1) sample, $\delta \geq 2M \sim 56.2$ μeV, $T$ =25 mK, and $k_BT$ =2.15 μeV. By setting $m_0$ =100, the transition occurs at $(\delta * m_0/m) \sim 2k_BT$=4.3 μeV. Therefore, the upper thickness limit of the QAH state is estimated to be $m \sim 1300$ (Supplementary Fig. 12). In other words, the QAH effect will disappear for $m \geq 1300$ because of the formation of the conducting gapless side surfaces.

To summarize, we realize the 3D QAH state in magnetic TI trilayers of hundred-nanometer thickness, exceeding ten times the thickest QAH sample record [1,3,4,7,19]. By varying magnetic dopant, gate voltage, temperature, and magnetic field, we demonstrate that the 3D QAH effect in our thick magnetic TI trilayers is robust. Our theoretical calculations show that the magnetization in the top and bottom surface layers can still gap the nonchiral side surface states in the hundred-nanometer-thick magnetic TI trilayers, resulting in the appearance of the 3D QAH state in these thick samples. Our study eliminates the thickness constraints traditionally associated with the QAH effect, expanding the scope of this concept to the 3D limit. The realization of the 3D QAH insulator of hundred-nanometer thickness also represents a very significant advance in MBE growth of topological materials and provides a platform for studying axion physics and image magnetic monopole[10] and exploring emergent magnetic topological phases such as higher-order TI.

**Methods**

**MBE growth**

All magnetic TI sandwiches, including 3 QL Cr-doped (Bi, Sb)$_2$Te$_3$/$m$ QL (Bi, Sb)$_2$Te$_3$/3 QL Cr-doped (Bi, Sb)$_2$Te$_3$ (designated as Cr-$m$) and 3 QL Cr/V-co-doped (Bi, Sb)$_2$Te$_3$/$m$ QL (Bi, Sb)$_2$Te$_3$/3 QL Cr/V-co-doped (Bi, Sb)$_2$Te$_3$ (designated as Cr/V-$m$), are grown in a commercial MBE system (Omicron Lab10) with a base vacuum better than ~2 × 10$^{-10}$ mbar. The heat-treated insulating SrTiO$_3$(111) substrates with a thickness of ~0.5 mm are first outgassed at ~600 °C for 1 hour before the growth of the magnetic TI sandwich samples. High-purity Bi (99.9999%), Sb (99.9999%), Cr (99.999%), V (99.999%), and Te (99.9999%) are evaporated from Knudsen effusion cells. During the growth of doped and undoped TI layers, the substrate is maintained at ~230 °C. The flux ratio of Te per (Bi + Sb + Cr/V) is set to be greater than ~10 to prevent Te deficiency in the films. The Bi/Sb ratio in each layer was optimized to tune the chemical potential of the entire magnetic TI sandwich near the charge neutral point. The growth rate of both magnetically doped TI and undoped TI films is ~0.2 QL per minute. No capping layer is involved in our *ex-situ* measurements.

**Electrical transport measurements**

All magnetic TI sandwich heterostructures grown on 2 mm × 10 mm insulating SrTiO$_3$(111) substrates are scratched into a Hall bar geometry using a computer-controlled probe station. The effective area of the Hall bar is ~1 mm × 0.5 mm. The electrical ohmic contacts are made by pressing indium dots onto the films. The bottom gate is prepared by flattening the indium dots on the back side of the SrTiO$_3$(111) substrates. Electrical transport measurements are conducted using both a Physical Property Measurement System (PPMS, Quantum Design DynaCool, 1.7K, 9T) for $T \geq 1.7$ K and a dilution refrigerator (Leiden Cryogenics, 10mK, 9T) for $T < 1.7$ K. The magnetic field is applied perpendicular to the sample plane. The bottom gate voltage $V_g$ is applied using a Keithley 2450 source meter. The PPMS measurements are carried out by an internal AC resistance bridge with an excitation current of ~1μA. The dilution measurements are carried out by a standard lock-in technique with an excitation current of ~1nA. The high magnetic field measurements are carried out with a superconducting magnet (SCM4, 32T, 50mK) and a hybrid magnet (Cell 15, 45T, 300mK) in NHMFL at Tallahassee. More transport results are found in Supplementary Figs. 4 to 11.

**Theoretical calculation**

We perform theoretical calculations of the side surface gap based on a sandwich model[45]:

$$\mathcal{H}(\mathbf{k}) = \begin{bmatrix} M(\mathbf{k}) & -iA_1 \partial_z & 0 & A_2 k_- \\ -iA_1 \partial_z & -M(\mathbf{k}) & A_2 k_- & 0 \\ 0 & A_2 k_+ & M(\mathbf{k}) & iA_1 \partial_z \\ A_2 k_+ & 0 & iA_1 \partial_z & -M(\mathbf{k}) \end{bmatrix} + H_X,$$

where $k_\pm = k_x \pm i k_y$, $M(\mathbf{k}) = M_0 + B_1 \partial_z^2 - B_2(k_x^2 + k_y^2)$, and $H_X = \Delta(z)\sigma_z \otimes \tau_0$. To simulate the QAH state, we discretize it into a tight-binding model along the $z$-axis between neighboring QL from $\mathcal{H}(\mathbf{k})$. We also assume the spatial-dependent exchange field $\Delta(z)$ takes the values $M$ in the top and bottom 3 QL magnetic TI layers and zero in the middle $m$ QL undoped TI layers, respectively. The parameters in our model are as follows: $M_0 = 0.28$eV, $A_1 = 2.2$eV · Å, $A_2 = 4.1$eV · Å, $B_1 = 10$eV · Å$^2$, $B_2 = 56.6$eV · Å$^2$ (Refs.[46,47]). The lattice constants are a = 4.14Å, c = 9.57Å (Refs.[46,47]). In our calculations, we set $\varepsilon_0(k)$ =0. Note that this simplification has no impact on the topology or the gap size near the Γ point. We choose the magnetic exchange gap $2M$ to be ~7 meV for our magnetic TI sandwiches. In our calculations, the side surface spectral function A($k_x$, $\omega$) is calculated from the recursive Green's function approach[48], which characterizes the density of states on the side surface (Fig. 5). We impose periodic boundary conditions along the *x*-direction with a good quantum number $k_x$, and calculate the semi-infinite lead surface Green's function $G_{1,1}$ along the *y*-direction to expose the side surface. By analyzing the side surface spectral function $A(k_x, \omega) = -\text{ImTr}(G_{1,1})$, we can determine the side surface energy gap δ.

**Acknowledgments:** We are grateful to Zhen Bi and Nitin Samarth for helpful discussions and to Kaya Wei, Elizabeth Green, Robert Nowell, and Ali Bangura for technical assistance in NHMFL at Tallahassee. This work is primarily supported by the NSF grant (DMR-2241327), including MBE growth, dilution transport measurements, and theoretical calculations. The sample characterization is supported by the ARO Award (W911NF2210159) and the Penn State MRSEC for Nanoscale Science (DMR-2011839). The PPMS measurements are supported by the DOE grant (DE-SC0023113). Work done at NHMFL is supported by NSF (DMR-2128556) and the State of Florida. C. -Z. C. acknowledges the support from the Gordon and Betty Moore Foundation's EPiQS Initiative (GBMF9063 to C. -Z. C.).

**Author contributions:** C. -Z. C. conceived and designed the experiment. Y. -F. Z., D. Z., and Z. -J. Y. grew all magnetic TI sandwich samples and fabricated the Hall bar devices. R. Z., L.-J. Z., and D. Z. performed the dilution measurements. Y. -F. Z., R. Z., and D. Z. performed the PPMS measurements. K. W. and H. Y. performed the STEM measurements. Z. -T. S., C. -X. L., and K. T. L. provided theoretical support. Y. -F. Z., R. Z., and C.-Z. C. analyzed the data and wrote the manuscript with inputs from all authors.

**Competing interests:** The authors declare no competing financial interests.

**Data availability:** The datasets generated during and/or analyzed during this study are available from the corresponding author upon reasonable request.

**Figures and figure captions:**

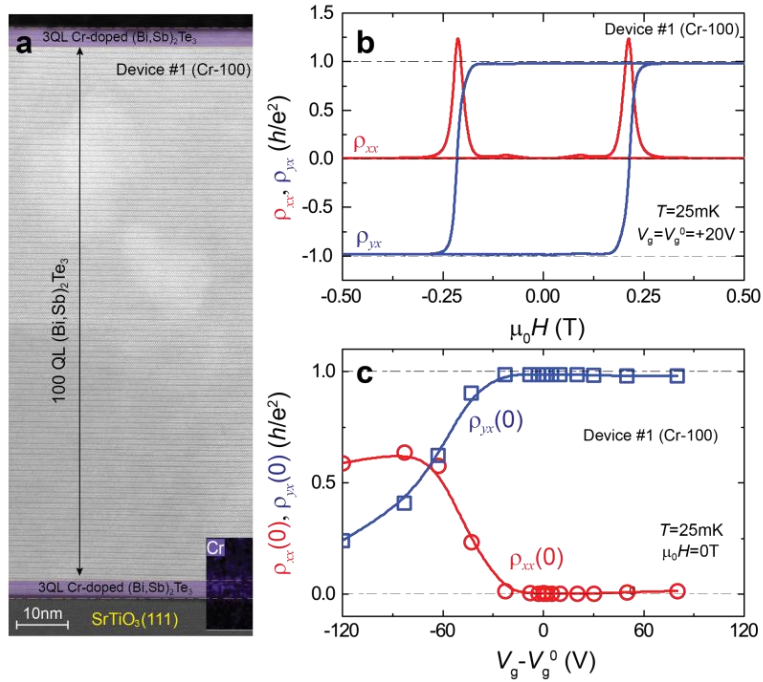

**Fig. 1| The QAH effect in MBE-grown 3 QL Cr-doped (Bi, Sb)$_2$Te$_3$/100 QL (Bi, Sb)$_2$Te$_3$/3 QL Cr-doped (Bi, Sb)$_2$Te$_3$ sandwich [Device #1 (Cr-100)]. a,** Cross-sectional STEM image. Inset: the EDS map of Cr near the bottom surface layers of the sample. **b,** Magnetic field $\mu_0 H$ dependence of the longitudinal resistance $\rho_{xx}$ (red) and the Hall resistance $\rho_{yx}$ (blue) at $V_g=V_g^0=+20$V and $T=25$mK. **c,** $(V_g-V_g^0)$ dependence of $\rho_{xx}(0)$ (red circle) and $\rho_{yx}(0)$ (blue square) at $T=25$mK and $\mu_0 H=0$T. $\rho_{xx}(0)$ and $\rho_{yx}(0)$ are the longitudinal resistance and Hall resistance under zero magnetic field, respectively.

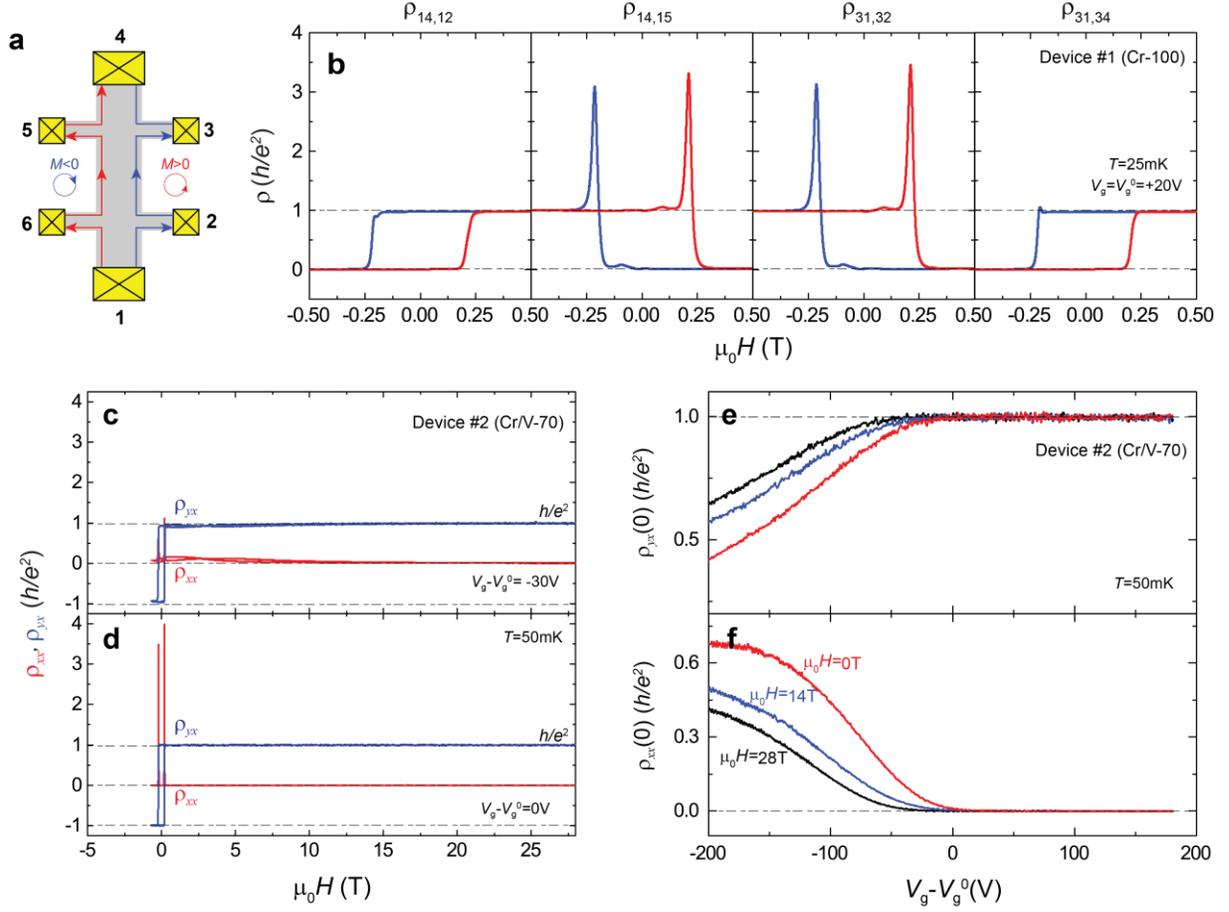

**Fig. 2| Demonstration of dissipation-free chiral edge transport in thick magnetic TI heterostructures [Device #1 (Cr-100) and Device #2 (Cr/V-70)]. a,** Schematics of chiral edge channels. The red and blue lines indicate the right- and left-handed chiral edge channels under positive and negative magnetizations, respectively. **b,** $\mu_0 H$ dependence of local and non-local three-terminal resistance $\rho_{14,12}$, $\rho_{14,13}$, $\rho_{31,32}$, and $\rho_{31,34}$ of Device #1 (Cr-100) at $V_g = V_g^0 = +20$ V and $T=25$mK. **c, d,** $\mu_0 H$ dependence of $\rho_{xx}$ (red) and $\rho_{yx}$ (blue) of Device #2 (Cr/V-70) at $T=25$mK and different $(V_g - V_g^0)$. (**c**) $(V_g - V_g^0) = -30$V and $(V_g - V_g^0)=0$V (**d**). **e, f,** $(V_g - V_g^0)$ dependence of $\rho_{yx}$ (**e**) and $\rho_{xx}$(**f**) of Device #2 (Cr/V-70) at $T=50$mK under different $\mu_0 H$. All the data shown here is not symmetrized.

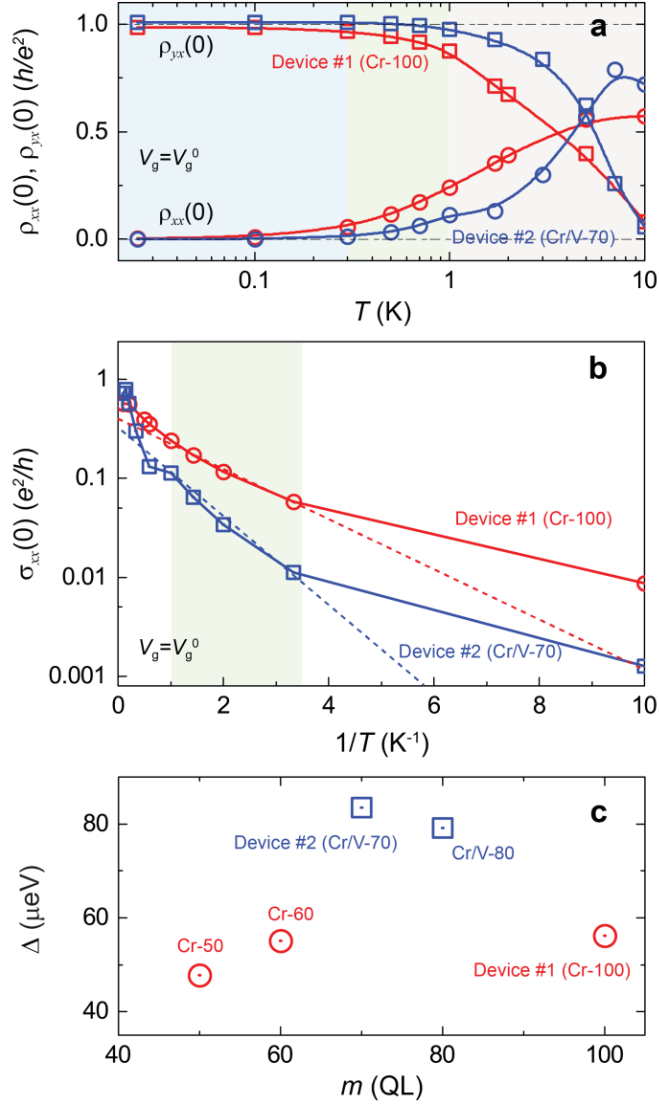

**Fig. 3| Comparation of the QAH states in thick Cr-doped and Cr/V-co-doped magnetic TI sandwiches. a,** Temperature dependence of $\rho_{xx}(0)$ (circle) and $\rho_{yx}(0)$ (square) in Device #1 (Cr-100) (red) and Device #2 (Cr/V-70) (blue). **b,** $\sigma_{xx}(0)$ as a function of $1/T$ in Device #1 (Cr-100) (red circle) and Device #2 (Cr/V-70) (blue square). The dashed lines show the fit of the Arrhenius function $\sigma_{xx} = \sigma_{xx}^0 e^{-\frac{E_a}{k_B T}}$. The fit temperature range is 0.3~1 K, which is shown in light green. **c,** $m$ dependence of the estimated thermal activation energy gap $\Delta$ in our five thick QAH sandwiches.

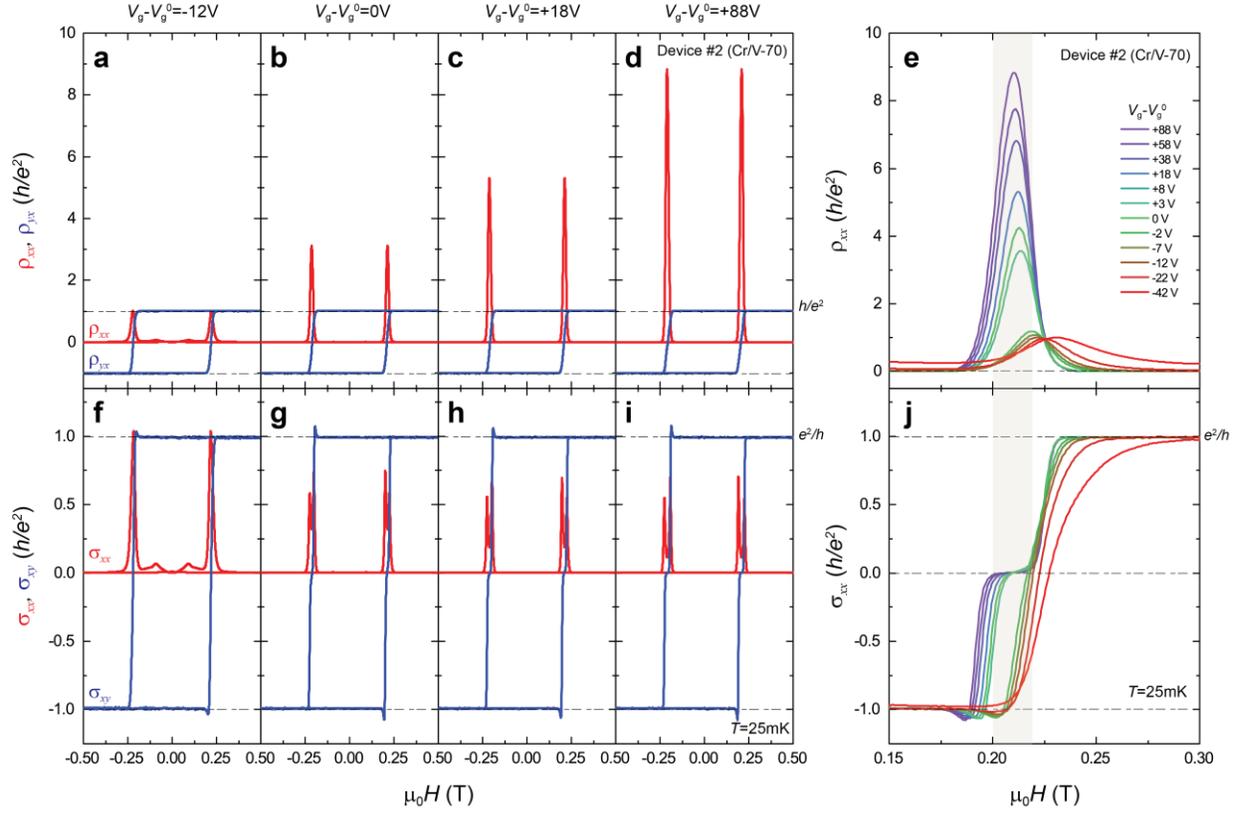

**Fig. 4| Gate-induced zero Hall conductance plateau in the Cr/V-co-doped TI sandwich [i.e., Device #2 (Cr/V-70)]. a-d,** $\mu_0H$ dependence of $\rho_{xx}$ (red) and $\rho_{yx}$ (blue) at $T$=25mK and different $(V_g-V_g^0)$. (**a**) $(V_g-V_g^0)$=-12V, (**b**) $(V_g-V_g^0)$=0V, (**c**) $(V_g-V_g^0)$= +18V, and (**d**) $(V_g-V_g^0)$=+88V. Note that $\rho_{xx}$ shows a change of 8.8 times while $\rho_{yx}$ stays quantized with the change in $V_g$. **e,** $\mu_0H$ dependence of $\rho_{xx}$ near coercive field $\mu_0H_c$ at $T$=25mK and different $(V_g-V_g^0)$. **f-i,** $\mu_0H$ dependence of $\sigma_{xx}$ (red) and $\sigma_{yx}$ (blue) at =25mK and different $(V_g-V_g^0)$. (**f**) $(V_g-V_g^0)$=-12V, (**j**) $(V_g-V_g^0)$=0V, (**h**) $(V_g-V_g^0)$= +18V, and (**i**) $(V_g-V_g^0)$=+88V. **j,** $\mu_0H$ dependence of $\sigma_{yx}$ near $\mu_0H_c$ at $T$=25mK and different $(V_g-V_g^0)$. The zero plateau regime is shown in grey.

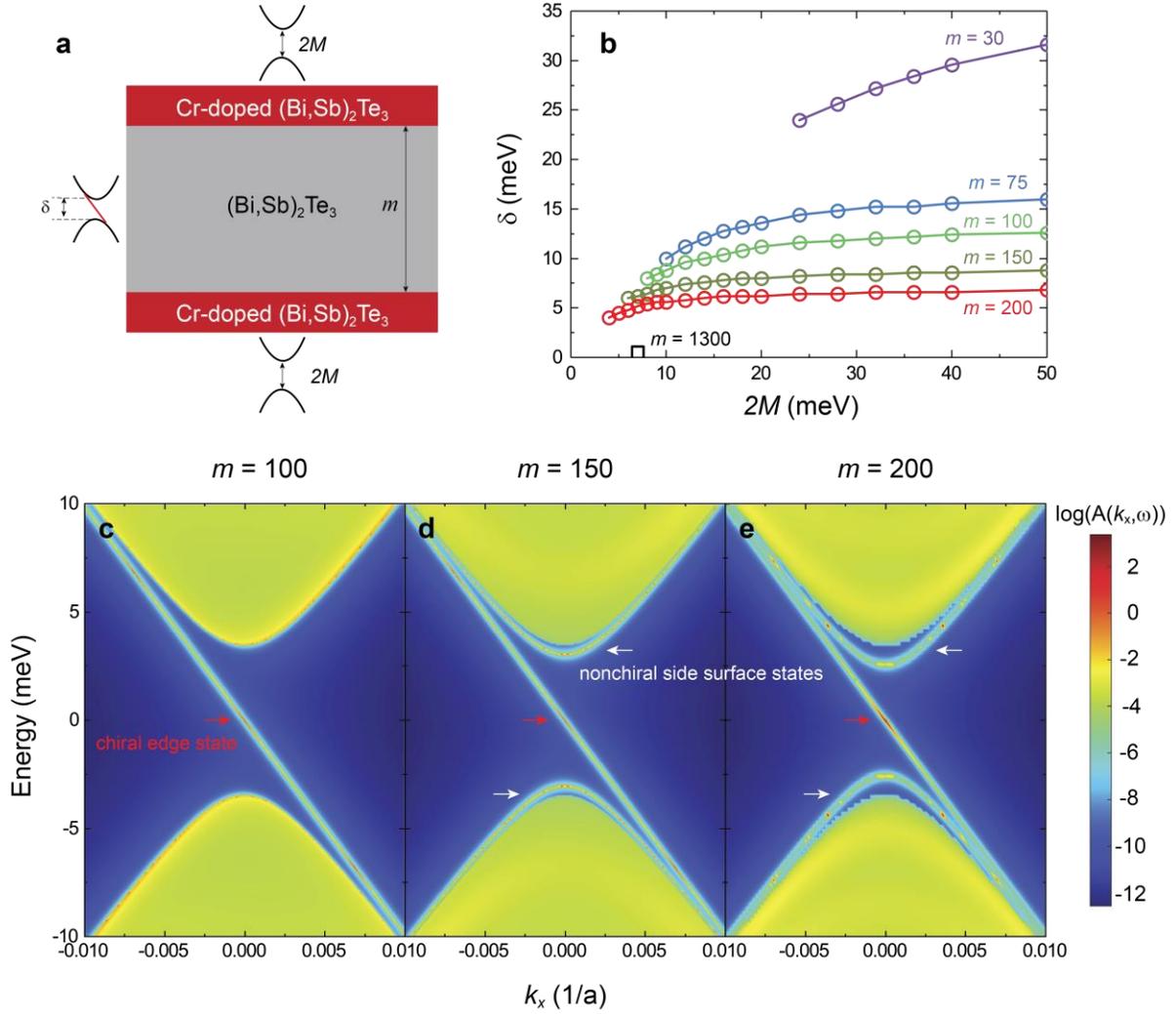

**Fig. 5| Quantum confinement-induced nonchiral side surface gaps in thick QAH insulators.**
**a,** Schematics of the magnetic TI trilayer. **b,** The nonchiral side surface gap $\delta$ as a function of $2M$ in QAH insulators with different $m$. **c-e**, Surface spectral functions in QAH insulators with $m =$ 100 (**c**), $m =$ 150 (**d**), $m =$ 200 (**e**). $2M = 7$ meV is used in (**c** to **e**). The chiral edge states and the gapped nonchiral side surface state are marked by red and white arrows, respectively.